\documentclass[aps,11pt,prd,superscriptaddress,preprint,nofootinbib]{revtex4}
\usepackage{amsmath}
\usepackage{graphics}
\usepackage{graphicx}
\usepackage{ae,aecompl}

\usepackage{bm}
\usepackage[bookmarks=true,bookmarksopen=true,bookmarksnumbered=true,bookmarksopenlevel=3]{hyperref}

\begin{document}
\newcommand{\bd}{\begin{document}}
\newcommand{\ed}{\end{document}}
\newcommand{\bc}{\begin{center}}
\newcommand{\ec}{\end{center}}
\newcommand{\bfr}{\begin{flushright}}
\newcommand{\efr}{\end{flushright}}
\newcommand{\lt}{\left}
\newcommand{\rt}{\right}
\newcommand{\vs}{\vspace}
\newcommand{\hs}{\hspace}
\newcommand{\beq}{\begin{equation}}
\newcommand{\eeq}{\end{equation}}
\newcommand{\lb}{\linebreak}
\newcommand{\pb}{\pagebreak}
\newcommand{\mb}{\makebox}
\newcommand{\fb}{\framebox}
\newcommand{\mc}{\multicolumn}
\newcommand{\ben}{\begin{enumerate}}
\newcommand{\een}{\end{enumerate}}
\newcommand{\bit}{\begin{itemize}}
\newcommand{\eit}{\end{itemize}}
\newcommand{\un}{\underline}
\newcommand{\lefq}{\lefteqn}
\newcommand{\ba}{\begin{array}}
\newcommand{\ea}{\end{array}}
\newcommand{\beqa}{\begin{eqnarray}}
\newcommand{\eeqa}{\end{eqnarray}}
\newcommand{\beqas}{\begin{eqnarray*}}
\newcommand{\eeqas}{\end{eqnarray*}}
\newcommand{\bfg}{\begin{figure}}
\newcommand{\efg}{\end{figure}}
\newcommand{\bds}{\begin{displaymath}}
\newcommand{\eds}{\end{displaymath}}
\newcommand{\btb}{\begin{tabbing}}
\newcommand{\etb}{\end{tabbing}}
\newcommand{\para}{\parallel}
\newcommand{\pad}{\partial}
\newcommand{\nn}{\nonumber}
\newcommand{\la}{\leftarrow}
\newcommand{\ra}{\rightarrow}
\newcommand{\lgla}{\longleftarrow}
\newcommand{\lgra}{\longrightarrow}
\newcommand{\La}{\Leftarrow}\newcommand{\Ra}{\Rightarrow}
\newcommand{\Lra}{\Leftrightarrow}
\newcommand{\Lgla}{\Longleftarrow}
\newcommand{\Lgra}{\Longrightarrow}
\newcommand{\lan}{\langle}
\newcommand{\ran}{\rangle}
\renewcommand{\a}{\alpha}
\renewcommand{\b}{\beta}
\newcommand{\g}{\gamma}
\newcommand{\G}{\Gamma}
\renewcommand{\d}{\delta}
\newcommand{\eps}{\epsilon}
\newcommand{\Th}{\Theta}
\newcommand{\s}{\sigma}
\newcommand{\lam}{\lambda}
\newcommand{\D}{\Delta}
\newcommand{\vare}{\varepsilon}
\newcommand{\pr}{\prime}
\newcommand{\ro}{\rho}
\newcommand{\nab}{\nabla}
\newcommand{\m}{\mu}
\newcommand{\n}{\nu}
\newcommand{\Sg}{\Sigma}
\newcommand{\p}{\pi}
\newcommand{\R}{I\!\!R}
\newcommand{\om}{\omega}
\newcommand{\Om}{\Omega}
\newcommand{\ze}{\zeta}
\newcommand{\vart}{\vartheta}
\newcommand{\tri}{\triangle}
\newcommand{\f}{\frac}
\newcommand{\iny}{\infty}
\newcommand{\pro}{\propto}
\title{Pseudo Hermitian interactions in the Dirac equation}

\author{\textsc{O.~Panella}}
\affiliation{Istituto Nazionale di Fisica Nucleare, Sezione di Perugia, Via A.~Pascoli, I-06123 Perugia, Italy}
\email{orlando.panella@pg.infn.it; Phone: +39 075 585 2762; Fax: +39 075 44666. ({\bf Corresponding Author}). }

\author{\textsc{P.~Roy}}
\affiliation{Physics and Applied Mathematics Unit\\
Indian Statistical Institute  \\ Kolkata,
India}

\begin{abstract}
We consider $(2+1)$ dimensional massless Dirac equation in the presence of complex vector potentials. It is shown that such vector potentials (leading to complex magnetic fields) can produce bound states and the Dirac Hamiltonians are $\eta$-pseudo Hermitian. Some examples have been explicitly worked out.
\end{abstract}


\maketitle

\section{Introduction}
In recent years massless Dirac equation in $(2+1)$ dimensions have drawn a lot of attention primarily because of its similarity to the equation governing the motion of charge carriers in graphene \cite{geim,geim1}. In view of the fact that electrostatic fields alone can not provide confinement of the electrons, there have been quite a number of works on exact solutions of the relevant Dirac equation with different magnetic field configurations, for example, square well magnetic barrier \cite{mag}, non zero magnetic fields in dots \cite{dot}, decaying magnetic fields \cite{ghosh}, solvable magnetic field configurations \cite{kuru} etc. On the other hand, at the same time there have been some investigations into the possible role of non Hermiticity and  $\cal{PT}$ symmetry \cite{bender} in graphene \cite{fagotti,saz,esaki}, optical analogues of relativistic quantum mechanics \cite{longhi} and relativistic non Hermitian quantum mechanics \cite{longhi1}, photonic honeycomb lattice \cite{ramezani} etc. Also $(2+1)$ dimensional Dirac equation with non Hermitian Rashba and scalar interaction was studied \cite{mandal}. Here our objective is to widen the scope of incorporating non Hermitian interactions in $(2+1)$ dimensional Dirac equation. We shall introduce non $\cal{PT}$ symmetric but non Hermitian interactions by using  imaginary vector potentials. It may be noted that imaginary vector potentials have been studied previously in connection with the localization/delocalization problem \cite{hatano,fein} as well as $\cal{PT}$ phase transition in higher dimensions \cite{yadav}. To be more specific, we shall consider $\eta$-pseudo Hermitian interactions \cite{mostafa} within the framework of $(2+1)$ dimensional massless Dirac equation. In particular, we shall examine exact bound state solutions in the presence of imaginary magnetic fields arising out of imaginary vector potentials. We shall also obtain the metric and it will be shown that the Dirac Hamiltonians are $\eta$-pseudo Hermitian. 

\section{The Model}
The motion of electrons in graphene is governed by an equation similar to the $(2+1)$ dimensional massless Dirac equation and is given by
\beq
H\psi=E\psi,~~~~~H=c{\bm{\sigma . P}}=c\left(\ba{cc} 0 & P_- \\ P_+ & 0 \ea\right)~,~~~~~\psi=\left(\ba{c}\psi_1 \\ \psi_2\ea\right)\label{eqn}
\eeq
where  
\beq 
P_{\pm}=(P_x\pm iP_y)=(p_x+A_x)\pm i(p_y+A_y)
\eeq
In order to solve Eq.(\ref{eqn}) it is necessary to decouple the spinor components. Applying the operator $H$ from the left in Eq.(\ref{eqn}) we find
\beq
c^2\left(\ba{cc}P_-P_+ & 0\\0 & P_+P_-\ea\right)\psi=E^2\psi\label{square}
\eeq
Let us now consider the vector potential to be
\beq
A_x=0,~~~~A_y=f(x)
\eeq
so that the magnetic field is given by
\beq
{\cal B}_z(x)=f^\prime(x)
\eeq
For the above choice of vector potentials the component wave functions can be taken of the form
\beq
\psi_{1,2}(x,y)=e^{ik_y y}\phi_{1,2}(x)
\eeq
Then from (\ref{square}) the equations for the components are found to be (in unit of $\hbar=1$)
\beq
\ba{l}
\displaystyle \left[-\f{d^2}{dx^2}+W^2(x)+W^\prime(x)\right]\phi_1(x)=\eps^2\phi_1(x)\\\\

\displaystyle \left[-\f{d^2}{dx^2}+W^2(x)-W^\prime(x)\right]\phi_2(x)=\eps^2\phi_2(x)\label{comp}\\\\
\ea
\eeq
where $\eps=(E/c)$ and the function $W(x)$ is given by
\beq
\displaystyle W(x)=k_y+f(x)
\eeq
\subsection{Complex Decaying Magnetic Field}
It is now necessary to choose the function $f(x)$. Our first choice for this function is 
\beq
f(x)=-(A+iB)~e^{-x},
\eeq
where $A>0$ and $B$ are constants. It leads to a complex exponentially decaying magnetic field
\beq
{\cal B}_z(x)=(A+iB)e^{-x}\label{mag1}
\eeq
For $B=0$ or a purely imaginary number (such that $(A+iB)>0$) the magnetic field is an exponentially decreasing one and we recover the case considered in ref \cite{ghosh,kuru}. 

Now from the second of Eq.(\ref{comp}) we obtain
\beq
\left[-\f{d^2}{dx^2}+V_2(x)\right]\phi_1=\left(\eps^2-k_y^2\right)\phi_1
\eeq
where 
\beq
V_2(x)=k_y^2+(A+iB)^2~e^{-2x}-(2k_y+1)(A+iB)~e^{-x}\label{pot2}
\eeq
It is not difficult to recognize $V_2(x)$ in Eq.(\ref{pot2}) as the complex analogue of the Morse potential whose solutions are well known \cite{flugge,cooper}. Using these results we find
\beq
\ba{l}
E_{2,n}=\pm c\sqrt{k_y^2-(k_y-n)^2}\\
\phi_{2,n}=t^{k_y-n}e^{-t/2}L_n^{(2k_y-2n)}(t),~~~~~n=0,1,2,....<[k_y]\label{sols1}
\ea
\eeq
where $t=2(A+iB)e^{-x}$ and $L_n^{(a)}(t)$ denotes generalized Laguerre polynomials. The first point to note here is that for the energy levels to be real it follows from (\ref{sols1}) that the corresponding eigenfunctions are acceptable when the condition $k_y\geq 0$ holds. For $k_y<0$, the wave functions are not normalizable i.e, no bound states are possible. 

Let us now examine the spin up component $\phi_1$. Since $\phi_2$ is known one can always use the intertwining relation 
\beq
c P_{-}\psi_{2}=E\psi_{1}\label{psi1}
\eeq
to obtain $\phi_1$. Nevertheless, for the sake completeness we present the explicit results for $\phi_1$. In this case the potential analogous to (\ref{pot2}) reads
\beq
V_1(x)=k_y^2+(A+iB)^2~e^{-2x}-(2k_y-1)(A+iB)~e^{-x}\label{pot1}
\eeq
Clearly, $V_1(x)$ can be obtained from $V_2(x)$ by the replacement $k_y\rightarrow k_y-1$ and so the solutions can be obtained from (\ref{sols1}) as
\beq
\ba{l}
E_{1,n}=\pm c\sqrt{k_y^2-(k_y-n-1)^2}\\
\phi_{1,n}=t^{k_y-n-1}e^{-t/2}L_n^{(2k_y-2n-2)}(t),~~~~~n=1,2,....<[k_y-1]\label{sols}
\ea
\eeq
Note that the $n=0$ state is missing from the spectrum (\ref{sols}) so that it is a spin down singlet state. Furthermore, $E_{2,n+1}=E_{1,n}$ so that the ground state is a singlet while the excited are doubly degenerate. Similarly the negative energy states are also paired. The precise structure of the wave functions are as follows (we present only the positive energy solutions):
\beq
\ba{l}
E_0=0,~~\psi_0=\left(\ba{cc} 0 \\ \phi_{2,0} \ea\right)\\
E_{n+1}= c\sqrt{k_y^2-(k_y-n-1)^2},~~\psi_{n+1}=\left(\ba{cc}\phi_{1,n}\\ \phi_{2,n+1} \ea\right),~~n=0,1,2,\cdots
\ea\eeq
It is interesting to note that the spectrum does not depend on the magnetic field. Also the dispersion relation is no longer linear as it should be in the presence of interactions. So, we find that it is indeed possible to create bound states with an imaginary vector potential. We shall now demonstrate the above results for a second example.

\subsection{Complex Hyperbolic Magnetic Field}
Here we choose $f(x)$ which leads to an effective potential of the complex hyperbolic Rosen-Morse type : 
\beq
f(x)=A~\text{tanh}(x-i\a),~~~~\text{$A$ and $\alpha$ are real constants}
\label{f2}
\eeq
In this case the complex magnetic field is given by
\beq
{\cal B}_z(x)=A~\text{sech}^2(x-i\a)\label{mag2}
\eeq
Note that for $\a=0$ we get back the results of refs \cite{kuru,mur}. Using (\ref{f2}) in second of Eq.(\ref{comp}) we find
\beq
\left[-\f{d^2}{dx^2}+U_2(x)\right]\phi_1=(\eps^2-k_y^2-A^2)\phi_1
\eeq
where 
\beq
U_2(x)=k_y^2-A(A+1)~\text{sech}^2(x-i\a)+2Ak_y~\text{tanh}(x-i\a)\label{pot2}
\eeq
This is the Hyperbolic Rosen-Morse potential with known energy values and eigenfunctions. In the present case the eigenvalues and the corresponding eigenfunctions are given by \cite{cooper,morse}
\beq
\ba{l}
E_{2,n}=\pm c\sqrt{A^2+k_y^2-(A-n)^2-\f{A^2k_y^2}{(A-n)^2}},~~n=0,1,2,....<[A-\sqrt{Ak_y}]\\\\
\displaystyle\phi_{2,n}=(1-t)^{s_1/2}(1+t)^{s_2/2}P_n^{(s_1,s_2)}(t)
\ea
\eeq
where $P_n^{(a,b)}(z)$ denotes Jacobi polynomials and 
\beq
t=\text{tanh}~x,~~s_{1,2}=A-n\pm \f{Ak_y}{A-n} 
\eeq
The energy values corresponding to the upper component of the spinor can be found out by replacing $A$ by $(A-1)$ and $\phi_1$ can be found out using the relation (\ref{psi1}).

\section{$\eta$-Pseudo Hermiticity}

It may be recalled that $\cal{PT}$ stands for parity inversion ($\cal{P}$) and time inversion ($\cal{T}$) and their actions are defined by \cite{bender}
\beq
\ba{l}
{\cal{P}} x {\cal{P}}=-x,~~~~{\cal{P}}p{\cal{P}}=-p\\\\

{\cal{T}} x{\cal{T}}=x,~~~~~{\cal{T}} p{\cal{T}} =-p,~~~~~{\cal{T}} i{\cal{T}}=-i
\ea
\eeq
So, a one dimensional non Hermitian Hamiltonian $H\neq H^\dag$ is $\cal{PT}$ symmetric if 
\beq
H{\cal{PT}}={\cal{PT}}H
\eeq
This last relation leads to the condition
\beq
V(x)=V^*(-x)\label{ptcond}
\eeq
It is easily seen that neither of the potentials $V_{1,2}(x)$ or $U_{1,2}$ above satisfy the condition (\ref{ptcond}) and therefore, are not $\cal{PT}$ symmetric. Nevertheless they admit real eigenvalues. To explain this feature, let us recall that a Hamiltonian is $\eta$-pseudo Hermitian if \cite{mostafa}
\beq
\eta H \eta^{-1}=H^\dag
\eeq
where $\eta$ is a Hermitian operator. It is known that eigenvalues of a $\eta$-pseudo Hermitian Hamiltonian are either all real or are complex conjugate pairs \cite{mostafa}. In view of the fact that in the present examples the eigenvalues are all real, one is tempted to conclude that the interactions are $\eta$ pseudo Hermitian. To this end we first consider case 1 and following ref \cite{ahmed}, let us consider the Hermitian operator
\beq
\eta=e^{-\theta p_x},~~~~\theta=\arctan\f{B}{A}
\eeq
Then it follows that
\beq
\eta c \eta^{-1}=c,~~~~\eta p_x\eta^{-1}=p_x,~~~~\eta V(x)\eta^{-1}=V(x+i\theta)\label{prop}
\eeq

We recall that in both the cases considered here the Hamiltonian is of the form
\beq
H=c\bm{\sigma . P}=c\left(\ba{cc} 0 & P_- \\ P_+ & 0 \ea\right)\label{dirachamil}
\eeq
where for the first example
\beq
P_{\pm}=p_x\pm ip_y\pm i(A+iB)e^{-x}\label{p}
\eeq
Then 
\beq
H^\dag=c\left(\ba{cc} 0 & P_+^\dag\\P_-^\dag & 0\ea\right)
\eeq
Now from Eq.(\ref{p}) it follows that 
\beq
P_+^\dag=p_x-ip_y-i(A-iB)e^{-x},~~~~P_-^\dag=p_x+ip_y+i(A-iB)e^{-x}
\eeq
and using Eq.(\ref{prop}) it can be shown that
\beq
\eta P_+\eta^{-1}=p_x+ip_y+i(A-iB)e^{-x}=P_-^\dag,~~~~\eta P_-\eta^{-1}=p_x-ip_y-i(A-iB)e^{-x}=P_+^\dag
\eeq
Next to demonstrate pseudo Hermiticity of the Dirac Hamiltonian (\ref{dirachamil}), let us consider the operator $\eta^\prime=\eta.{\cal I}_2$ where ${\cal I}_2$ is the $(2\times 2)$ unit matrix. Then it can be shown that
\beq
\eta^\prime H {\eta^\prime}^{-1}=H^\dag
\eeq
Thus the Dirac Hamiltonian with a complex decaying magnetic field (\ref{mag1}) is $\eta$-pseudo Hermitian. 

For the magnetic field given by (\ref{mag2}) the metric operator $\eta$ can be found by using the relations (\ref{prop}). After a straightforward calculation it can be shown that the metric operator is given by
\beq
\eta=e^{-2\a p_x}
\eeq
so that in this second example also the Dirac Hamiltonian is $\eta$-pseudo Hermitian.

\section{Conclusions}
Here we have studied $(2+1)$ dimensional massless Dirac equation~\footnote{We note that if a massive particle of mass $m$ is considered, the energy spectrum in the first example would become $E_n=c\sqrt{k_y^2+m^2c^2-(k_y-n)^2}$. Similar changes will occur in the second example too.} in the presence of complex magnetic fields and it has been shown that such magnetic fields can create bound states. 
It has also been shown that Dirac Hamiltonians in the presence of such magnetic fields are $\eta$-pseudo Hermitian. We feel it would be of interest to study generation of bound states using other types of magnetic fields e.g, periodic magnetic fields.

\begin{acknowledgments}
One of us (P.~R.) wishes to thank INFN Sezione di Perugia for supporting a visit during which part of this work was carried out.  He would also like to thank the Physics Department of the University of Perugia
for hospitality.
\end{acknowledgments}

\ed